# Secure and authenticated quantum secret sharing

*Muhammad Nadeem*
Department of Basic Sciences,
School of Electrical Engineering and Computer Science,
National University of Sciences and Technology (NUST),
H-12 Islamabad, Pakistan
muhammad.nadeem@seecs.edu.pk

*Noor Ul Ain*
Department of Computer Sciences,
School of Electrical Engineering and Computer Science,
National University of Sciences and Technology (NUST),
H-12 Islamabad, Pakistan
13msccsnaain@seecs.nust.edu.pk

We propose here a quantum secret sharing scheme that works for both quantum and classical secrets. The proposed scheme is based on both entanglement swapping and teleportation together. It allows sender to encrypt his/her secret and simultaneously distribute decryption information of the encrypted secret among distant parties where decryption shares are non-locally correlated with each other. Two-fold quantum non-local correlations, generated through entanglement swapping and then teleporting quantum states over swapped maximally entangled pairs, guarantee both authentication and secrecy of the secret from internal as well as external eavesdroppers. For classical secrets, we demonstrate a direct (2,2) quantum secret sharing scheme where neither pre-shared key nor physically secure quantum/classical channel are required. The same scheme turns out to be a (5,5) quantum secret sharing scheme for quantum secrets if sender and receivers have private quantum/classical channels among them.

**Introduction**
Secret sharing is a cryptographic procedure where a secret (or decryption function of secret) is divided among multiple parties such that each party gets a part of secret (or decryption function of secret) but cannot extract the secret on its own. However, the secret can only be reconstructed (extracted) when all the parties collaborate together and share their part. Such a secret sharing scheme is called $(n, n)$ threshold scheme. In general, the original secret can be reconstructed if and only if at least $k \leq n$ parties collaborate called a $(k, n)$ secret sharing threshold scheme.

Secret sharing is an important cryptographic primitive that can be used as sub-protocol for implementing secure multiparty computation, multiparty key agreement, hardware security modules, PGP key recovery, and visual cryptography. The idea of classical secret sharing was given originally by Shamir and Blakely in 1979 [1,2] and was extended to the quantum regime by M. Hillery et al [3] in 1999. Quantum secret sharing scheme proposed by M. Hillery was based on three-particle GHZ states. Later on, Karlsson *et al* showed that quantum secret sharing protocol similar to that of M. Hillery can also be constructed by using two-particle quantum entangled states [4]. After generalization of secret sharing to quantum regime [3,4], a number of quantum secret sharing schemes are developed for both classical (bits) as well as quantum (qubits) secrets [5-27].

Here we show that, by using two-particle quantum entangled states, a secure and authenticated (2,2) quantum secret sharing can be constructed without having any pre-shared data or private quantum/classical channels between sender and receivers. The proposed scheme is based on two-fold quantum non-local correlations generated through fundamental and important quantum information tasks namely entanglement swapping [28] and teleportation [29] together. By using at least two EPR pairs with receivers $R_1$ and $R_2$, two-fold quantum non-local correlations allow sender to encrypt his/her classical secret $|\xi\rangle$ (either $|\xi\rangle=|0\rangle$ or $|\xi\rangle=|1\rangle$) and distribute non-locally correlated decryption information between receivers $R_1$ and $R_2$ such that either $R_1$ or $R_2$ cannot decrypt the secret alone but they can find original secret if and only if they meet and share distributed decryption information.

The proposed procedure for secret sharing guarantees secrecy from both internal (receivers $R_1$ and $R_2$) as well as external (third party) eavesdroppers. It allows sender as well as receivers to perform repetitive measurements in publically known basis and extract authentication token. Repetitive measurements allow both sender and receivers to store classical information and later extract encrypted secret after indefinite time. After verifying authentication tokens from both receivers with quantum non-local correlations, sender reveals his/her share of decryption information publically.

The proposed (2,2) secret sharing scheme for classical secret can be used as a (5,5) secret sharing scheme for quantum secrets $|\xi\rangle = r|0\rangle + s|1\rangle$. In that case, sender needs to have either physically secure quantum/classical channels with receivers or they can establish secure channels through QKD or EPR based quantum channels.

Such a quantum procedure based on entanglement swapping and teleportation together can be used to achieve a number of cryptographic applications such as unconditionally secure bit commitment with arbitrarily long commitment time [30,31], two-sided two-party secure computation, asynchronous ideal coin tossing [32], and secure positioning [33-35].

**Teleportation and entanglement swapping**

Suppose sender S and receiver R share one of the four EPR pairs

$$|\Phi^+\rangle = (|00\rangle + |11\rangle)/\sqrt{2} \quad (1)$$

$$|\Psi^+\rangle = (|01\rangle + |10\rangle)/\sqrt{2} \quad (2)$$

$$|\Phi^-\rangle = (|00\rangle - |11\rangle)/\sqrt{2} \quad (3)$$

$$|\Psi^-\rangle = (|01\rangle - |10\rangle)/\sqrt{2} \quad (4)$$

and S wants to send an arbitrary quantum state $|\xi\rangle$ to R through teleportation [29]. He performs Bell state measurement (BSM) [36] on quantum state $|\xi\rangle$ and his half of EPR pair such that both quantum state and EPR pair collapsed and he gets classical 2-bit string $ss'$. On the other hand, entangled half in possession of R becomes $|\xi'\rangle = \tau_i |\xi\rangle$ where teleportation encoding $\tau_i$ depends upon both classical string $ss'$ and identity of shared EPR pair. In other word, receiver R can decode quantum state $|\xi\rangle$ from $|\xi'\rangle = \tau_i |\xi\rangle$ if and only if he knows both classical string $ss'$ and identity of shared EPR pair as shown in the table 1.

Entanglement swapping [28] is an interesting extension of teleportation [29], in fact, teleportation of entanglement, where two uncorrelated quantum systems gets entangled without having any interaction with each other. Let's consider three distant parties S, $R_1$, and $R_2$ where $R_1$ shares EPR pairs $|.\rangle_{sr_1}$ and $|.\rangle_{r_1 r_2}$ with S and $R_2$ respectively while S and $R_2$ are uncorrelated. Now if $R_1$ performs BSM on his/her halves, quantum systems in possession of S and $R_2$ gets entangled into one of the four possible Bell states $|.\rangle_{sr_2}$. Exact identity if EPR pair $|.\rangle_{sr_2}$ swapped between S and $R_2$ depends upon both initially shared EPR pairs $|.\rangle_{sr_1}$ and $|.\rangle_{r_1 r_2}$ as well as BSM result of $R_1$ $|.\rangle_{r_1 r_1}$. All possible transformations of initially shared quantum system $|.\rangle_{sr_1}|.\rangle_{r_1 r_2}$ to $|.\rangle_{r_1 r_1}|.\rangle_{sr_2}$ are summarized in table 2.

| EPR | $ss'$ | | | | $|\{'\rangle = \dagger_i|\{\rangle$ | | | |
|---|---|---|---|---|---|---|---|---|
| $|\Phi^+\rangle$ | 00 | 01 | 10 | 11 | $|\{\rangle$ | $\dagger_x|\{\rangle$ | $\dagger_z|\{\rangle$ | $\dagger_z\dagger_x|\{\rangle$ |
| $|\Psi^+\rangle$ | 00 | 01 | 10 | 11 | $\dagger_x|\{\rangle$ | $|\{\rangle$ | $\dagger_z\dagger_x|\{\rangle$ | $\dagger_z|\{\rangle$ |
| $|\Phi^-\rangle$ | 00 | 01 | 10 | 11 | $\dagger_z|\{\rangle$ | $\dagger_z\dagger_x|\{\rangle$ | $|\{\rangle$ | $\dagger_x|\{\rangle$ |
| $|\Psi^-\rangle$ | 00 | 01 | 10 | 11 | $\dagger_z\dagger_x|\{\rangle$ | $\dagger_z|\{\rangle$ | $\dagger_x|\{\rangle$ | $|\{\rangle$ |

**Table 1:** Teleportation: If sender S and receiver R share entangled state $|\Phi^-\rangle$ and BSM result of S is $ss' = 11$ then R will have state $|\{'\rangle = \dagger_x|\{\rangle$ on his side.

| $|.\rangle_{sr_1}|.\rangle_{r_1 r_2}$ | | | | | | | | $|.\rangle_{r_1 r_1}|.\rangle_{sr_2}$ | | | | | | | |
|---|---|---|---|---|---|---|---|---|---|---|---|---|---|---|---|
| $|\Phi^+\rangle|\Phi^+\rangle$ | | $|\Psi^+\rangle|\Psi^+\rangle$ | | $|\Phi^-\rangle|\Phi^-\rangle$ | | $|\Psi^-\rangle|\Psi^-\rangle$ | | $|\Phi^+\rangle|\Phi^+\rangle$ | | $|\Psi^+\rangle|\Psi^+\rangle$ | | $|\Phi^-\rangle|\Phi^-\rangle$ | | $|\Psi^-\rangle|\Psi^-\rangle$ | |
| $|\Phi^+\rangle|\Psi^+\rangle$ | | $|\Psi^+\rangle|\Phi^+\rangle$ | | $|\Phi^-\rangle|\Psi^-\rangle$ | | $|\Psi^-\rangle|\Phi^-\rangle$ | | $|\Phi^+\rangle|\Psi^+\rangle$ | | $|\Psi^+\rangle|\Phi^+\rangle$ | | $|\Phi^-\rangle|\Psi^-\rangle$ | | $|\Psi^-\rangle|\Phi^-\rangle$ | |
| $|\Phi^+\rangle|\Phi^-\rangle$ | | $|\Psi^+\rangle|\Psi^-\rangle$ | | $|\Phi^-\rangle|\Phi^+\rangle$ | | $|\Psi^-\rangle|\Psi^+\rangle$ | | $|\Phi^+\rangle|\Phi^-\rangle$ | | $|\Psi^+\rangle|\Psi^-\rangle$ | | $|\Phi^-\rangle|\Phi^+\rangle$ | | $|\Psi^-\rangle|\Psi^+\rangle$ | |
| $|\Phi^+\rangle|\Psi^-\rangle$ | | $|\Psi^+\rangle|\Phi^-\rangle$ | | $|\Phi^-\rangle|\Psi^+\rangle$ | | $|\Psi^-\rangle|\Phi^+\rangle$ | | $|\Phi^+\rangle|\Psi^-\rangle$ | | $|\Psi^+\rangle|\Phi^-\rangle$ | | $|\Phi^-\rangle|\Psi^+\rangle$ | | $|\Psi^-\rangle|\Phi^+\rangle$ | |

**Table 2:** Entanglement swapping**:** After BSM of $R_1$, initially shared quantum system $|.\rangle_{sr_1}|.\rangle_{r_1 r_2} = |\Phi^+\rangle|\Psi^-\rangle$, could be transformed to one of the four possibilities, $|.\rangle_{r_1 r_1}|.\rangle_{sr_2} = |\Phi^+\rangle|\Psi^-\rangle$, $|.\rangle_{r_1 r_1}|.\rangle_{sr_2} = |\Psi^+\rangle|\Phi^-\rangle$, $|.\rangle_{r_1 r_1}|.\rangle_{sr_2} = |\Phi^-\rangle|\Psi^+\rangle$, or $|.\rangle_{r_1 r_1}|.\rangle_{sr_2} = |\Psi^-\rangle|\Phi^+\rangle$.

**Two-fold quantum non-local correlations and information splitting**
Here we show that when entanglement swapping and teleportation, as discussed above, are performed together on a single quantum system consisting of at least two EPR pairs, it results in perfect encryption of a secret (quantum/classical) with non-locally correlated classical shares of decryption function. Both these cryptographic requirements, perfect encryption along with set of decryption pieces, directly lead to secure and authenticated quantum secret sharing schemes.

Suppose sender S secretly prepares two maximally entangled quantum systems $H_{r_1 r'_1} = H_{r_1} \otimes H_{r'_1}$ and $H_{r_2 r'_2} = H_{r_2} \otimes H_{r'_2}$ in one of the Bell basis $|\Phi^+\rangle$, $|\Psi^+\rangle$, $|\Phi^-\rangle$, or $|\Psi^-\rangle$ corresponding to 2-bit strings $r_i r'_i = 00$, $r_i r'_i = 01$, $r_i r'_i = 10$, or $r_i r'_i = 11$ respectively. He then sends second half of each system to receiver $R_1$ while first half of second system to receiver $R_2$. As a result, sender S shares quantum system $H_{r_1} \otimes H_{r_1'} \otimes H_{r_2'} \otimes H_{r_2}$ with two receivers $R_1$ and $R_2$ where $H_{r_1}$ is kept by S while $H_{r_1'} \otimes H_{r_2'}$ and $H_{r_2}$ are in possession of $R_1$ and $R_2$ respectively.

If $R_1$ applies Bell operator [36] on $H_{r_1'}$ and $H_{r_2'}$ while S applies Bell operator on his secret $H$ and retained half $H_{r_1}$, then $R_2$'s half $H_{r_2}$ becomes $H_{r_2} = U(H)$ where $U$ is a $2 \times 2$ unitary operator. The unitary encoding $U$ is non-locally correlated with classical Bell state measurement (BSM) results $R_1 R_1' \in \{00, 01, 10, 11\}$ and $ss' \in \{00, 01, 10, 11\}$ of receiver $R_1$ and sender S respectively as well as identity of initially shared systems $r_1 r_1' \in \{00, 01, 10, 11\}$ and $r_2 r_2' \in \{00, 01, 10, 11\}$.

In other words, decoding information of secret $H$ is divided into five pieces; four equal shares of classical 2-bit strings $r_1 r_1'$, $r_2 r_2'$, $R_1 R_1'$ and $ss'$, and one encrypted secret $H_{r_2} = U(H)$. We would like to highlight here that for each value of $R_1$'s BSM result $R_1 R_1'$ obtained through Bell measurement on $H_{r_1'} \otimes H_{r_2'}$, there will be a unique Bell system $H_{r_1} \otimes H_{r_2}$ swapped between S and $R_2$ and hence unique teleportation encoding of quantum system $H$ corresponding to BSM result of Bob. As a result, secret $H$ can be decoded from $H_{r_2} = U(H)$ if and only if someone has four shares of decrypting information $r_1 r_1'$, $r_2 r_2'$, $R_1 R_1'$ and $ss'$ at same location [18].

**Quantum secret sharing - classical information**
Let's consider first a more general situation where sender S and two receivers $R_1$ and $R_2$ are distant parties and they have no pre-shared secret data among them. Our scheme guarantees secure and authenticated sharing of classical secret information between two receivers under standard quantum cryptographic requirements [37-39]: (i) Quantum transmissions can be actively intercepted. (ii) Classical channels can be monitored passively. (iii) However, classical communication cannot be altered or suppressed. Detailed (2,2) quantum secret sharing scheme for classical information is described below.

**Authentication tokens:** Sender S shares two publically known EPR pairs with each of receivers $R_1$ and $R_2$, say $|\Phi^+\rangle$ and $|\Phi^-\rangle$ with receiver $R_1$ while $|\Psi^+\rangle$ and $|\Psi^-\rangle$ with receiver $R_2$. Both $R_1$ and $R_2$ perform BSM on their halves and keep their respective BSM results $r_1 r_1'$ and $r_2 r_2'$ secret from each other. However, their BSM results are known to sender S; S stores 2-bit strings $r_1 r_1'$ and $r_2 r_2'$ corresponding to his BSM on retained pairs.

**Information splitting between $R_1$ and $R_2$:** In the second round, S prepares two EPR pairs $H_{r_1 r_1'} = H_{r_1} \otimes H_{r_1'}$ and $H_{r_2 r_2'} = H_{r_2} \otimes H_{r_2'}$ corresponding to BSM results of $R_1$ and $R_2$ respectively. He then sends $H_{r_1'}$ and $H_{r_2'}$ to receiver $R_1$ while $H_{r_2}$ to receiver $R_2$. Receiver $R_1$ performs BSM on $H_{r_1'}$ and $H_{r_2'}$ and gets classical 2-bit string $R_1 R_1'$ while sender S and receiver $R_2$ shares entangled system $H_{r_1} \otimes H_{r_2}$. The sender S teleports his secret $|\varsigma\rangle \in \{0,1\}$ to $R_2$ over EPR channel $H_{r_1} \otimes H_{r_2}$ and keeps his BSM result $ss' \in \{00, 01, 10, 11\}$ secret. $R_2$ measures his half, that has been transformed to $|\varsigma'\rangle = \dagger_t |\varsigma\rangle$ now where $\dagger_t$ is teleportation encoding, and gets an arbitrary outcome $\varsigma' \in \{0,1\}$.

**Authentication:** $R_1$ bit-wise XOR his BSM results $r_1 r_1'$ and $R_1 R_1'$ and sends $(r_1 \oplus R_1)(r_1' \oplus R_1')$ to S over classical channel. Similarly, $R_2$ XOR measurement outcome $\varsigma' \in \{0,1\}$ with $r_2 \oplus r_2'$ and sends $\{' \oplus r_2 \oplus r'_2\}$ to S. The sender S extracts $R_1 R'_1$ through bit-wise XOR of $(r_1 \oplus R_1)(r'_1 \oplus R'_1)$ and $r_1 r'_1$ while $\{'$ through XOR of $\{' \oplus r_2 \oplus r'_2$ and $r_2 \oplus r'_2$. If sender S finds $R_1 R'_1$ and $\{'$ consistent with his secret $|\{\rangle$, he sends his BSM result $ss'$ to both $R_1$ and $R_2$ over public channels.

**Combining secret shares:** Receivers $R_1$ and $R_2$ can extract encoded message $|\{\rangle$ from $|\{'\rangle = \dagger_t |\{\rangle$ only if they meet and share their secrets: $r_1 r'_1$ and $R_1 R'_1$ kept by $R_1$ while $\{'$ and $r_2 r'_2$ kept by $R_2$ respectively.

**Quantum secret sharing - quantum information**
Now let's consider sender S has five distant agents $R_1, R_2, .., R_5$ such that sender S can securely communicate both quantum and classical information with his agents; either they have physically secure quantum/classical channels between them or they can establish secure quantum channel through QKD or EPR based quantum channels. In this situation, our proposed (2,2) quantum secret sharing scheme can be used as (5,5) quantum secret sharing scheme for quantum information $|\{\rangle = r|0\rangle + s|1\rangle$ without requiring authentication.

**Information splitting:** S prepares two EPR pairs $H_{s_1 s'_1} = H_{s_1} \otimes H_{s'_1}$ and $H_{s_2 s'_2} = H_{s_2} \otimes H_{s'_2}$ and sends $H_{s'_1}$ and $H_{s'_2}$ to receiver $R_1$ while $H_{s_2}$ to receiver $R_2$. Simultaneously, he sends 2-bit strings $s_1 s'_1$ and $s_2 s'_2$ to receivers $R_3$ and $R_4$ respectively. Receiver $R_1$ performs BSM on $H_{s'_1}$ and $H_{s'_2}$ and gets classical 2-bit string $R_1 R'_1$ while sender S and receiver $R_2$ shares entangled system $H_{s_1} \otimes H_{s_2}$. The sender S teleports his secret $|\{\rangle = r|0\rangle + s|1\rangle$ to $R_2$ and sends his BSM result $ss' \in \{00, 01, 10, 11\}$ to receiver $R_5$. At the end of this phase, $R_2$'s half has been transformed to $|\{'\rangle = \dagger_t |\{\rangle$ where $\dagger_t$ is teleportation encoding non-locally correlated with four equal shares of classical information $s_1 s'_1$, $s_2 s'_2$, $ss'$, and $R_1 R'_1$.

**Decoding secret:** Receivers $R_1, R_2, .., R_5$ can extract encoded message $|\{\rangle$ from $|\{'\rangle = \dagger_t |\{\rangle$ only if they meet and share their secrets $R_1 R'_1$, $|\{'\rangle = \dagger_t |\{\rangle$, $s_1 s'_1$, $s_2 s'_2$ and $ss'$ respectively.

**Discussion**
We proposed here a quantum secret sharing scheme based on two-fold quantum non-local correlations that assure authentication and secrecy of secret from internal as well as external eavesdropping simultaneously. The proposed quantum secret sharing scheme can be used as (2,2) threshold scheme for classical secret where both authenticity and secrecy is guaranteed without relying on advanced quantum technology, pre-shared secret keys or private quantum/classical channels between distant users. Repetitive measurements and classical communication over public channels assures availability of secret information and result in secure and authenticated quantum secret sharing scheme.

Only requirement for security/availability of secret information against eavesdroppers/noise is unsuppressed classical communication over public channels between distant users. The proposed procedure then remains secure against passive monitoring of classical information as well as active quantum attacks. It does not use batches of encoded qubits and then statistical tests after measurements in agreed basis and hence does not allow eavesdroppers to successfully use quantum attacks such as entangling a quantum ancilla with the encrypted message and later performing a specific measurement on it according to public communication of legitimate users. However, if eavesdroppers can interrupt classical communication actively then distant users need to have some pre-agreed secret key or trusted source for secure distribution of EPR pairs.

Finally, our proposed (2,2) quantum secret sharing scheme for classical secrets can be used as (5,5) quantum secret sharing scheme for quantum secret (qubit). For (5,5) threshold scheme, sender needs to have either physically secure quantum/classical channels with the receivers or they can establish secure quantum channel through QKD or EPR based quantum channels. Since quantum information, in the form of an arbitrary qubit $|\{\rangle = r|0\rangle + s|1\rangle$, is destroyed after measurement in computational basis, a similar approach for authentication as discussed for classical secret is not possible.